\documentclass[pra]{revtex4}
\newtheorem{md}{Theorem}
\newtheorem{lem1}[md]{Lemma}
\newtheorem{gersgorin}[md]{Theorem}
\newtheorem{onlyU}[md]{Theorem}
\usepackage{graphicx}
\usepackage{amssymb}

\begin{document}
\title{Phase boundaries in deterministic dense coding}
\author{Michael R. Beran$^1$ and Scott M. Cohen$^{1,2}$}
\email{cohensm@duq.edu}
\affiliation{$^1$Department of Physics, Duquesne University, Pittsburgh,
Pennsylvania 15282\\$^2$Department of Physics, Carnegie-Mellon University,
Pittsburgh, Pennsylvania 15213
}

\begin{abstract}
We consider dense coding with partially entangled states on bipartite systems of dimension $d\times d$, studying the conditions under which a given number of messages, $N$, can be deterministically transmitted. It is known that the largest Schmidt coefficient, $\lambda_0$, must obey the bound $\lambda_0\le d/N$, and considerable empirical evidence points to the conclusion that there exist states satisfying $\lambda_0=d/N$ for every $d$ and $N$ except the special cases $N=d+1$ and $N=d^2-1$. We provide additional conditions under which this bound cannot be reached -- that is, when it must be that $\lambda_0<d/N$ -- yielding insight into the shapes of boundaries separating entangled states that allow $N$ messages from those that allow only $N-1$. We also show that these conclusions hold no matter what operations are used for the encoding, and in so doing, identify circumstances under which unitary encoding is strictly better than non-unitary.
\end{abstract}

\maketitle
\section{Introduction}
Dense coding, which utilizes entangled quantum states to increase the classical communication capacity of a quantum channel, provided an early impetus for the recent growth of interest in the exciting field of quantum information. The first demonstration of this effect \cite{BennettDense} involved two spatially separated parties sharing a bipartite quantum system of dimension $d\times d$ in a maximally entangled state. Alice locally performs a unitary operation to encode her message, sends her part of the shared system through a noiseless quantum channel to Bob, who is then able to determine the message with certainty using a projective measurement on the combined parts. Since the original discovery, numerous variations of this protocol have been introduced and studied, including what is known as deterministic dense coding, where the messages must still always be identified with certainty, but the shared state may now be less than maximally entangled.

Deterministic dense coding with non-maximally entangled states was first studied in \cite{Mozes} and later in \cite{ourDense,Ji,GerjuoyDense,BeranK}. The main thrust of these papers was to study the ``maximal alphabet" (maximum number of messages, N) that can be deterministically transmitted using a given partially entangled state. An important observation that arose from the work of \cite{Mozes} and was shown there to be necessary when $N$ is an integer multiple of the dimension $d$, was that the largest Schmidt coefficient, $\lambda_0$, of the entangled state appeared to obey the bound $\lambda_0\le d/N$. Their observations also seemed to indicate that every boundary saturated that bound (intersected the plane $\lambda_0=d/N$) somewhere, except for the special cases of $N=d+1$ and $N=d^2-1$. This bound was later proved in \cite{ourDense} to be necessary for all values of $N$ and $d$, even if Alice is allowed to use the most general quantum operations to encode her messages. It was later proved that this bound could not be saturated for the cases $N=d+1$ \cite{GerjuoyDense} and $N=d^2-1$ \cite{Ji}. 

These results provide principles aiding the determination of the location of boundaries between what we will refer to as ``phases": regions in the parameter space of Schmidt coefficients within which a given number of messages can be sent. Nonetheless, establishing the positions and shapes of these boundaries remains a difficult task. In the present paper, we present numerical calculations determining the full phase diagram for $d=4$. We then prove two theorems, inspired by our $d=4$ results but valid for all $d$, that aid in understanding the shape of these phase boundaries.

After briefly reviewing the detailed setup of these protocols in the following section, and presenting our numerically generated phase diagram for $d=4$ in Section~\ref{sec:num}, we then turn to our main results in the subsequent sections. The first theorem is presented in Section~\ref{sec:shapes}. For a certain class of boundaries and when Alice is restricted to using unitary operations to encode her messages, this theorem provides conditions under which $\lambda_0$ must be strictly less than the bound $d/N$, and applies for all dimensions $d$. Given the strong empirical evidence that every boundary saturates this bound somewhere (apart from the exceptions noted above), we see that these boundaries curve away from the surface, $\lambda_0=d/N$, toward a (hyper-)plane, which is identified in the theorem (this theorem also shows that the bound cannot be reached anywhere on the boundary corresponding to $N=d+1$, generalizing the original proof of \cite{GerjuoyDense}). Thus, we have a general principle determining, in part, the shape (or at least the orientation) of these boundaries. Then, in Section~\ref{sec:onlyU}, we show that the conclusion of the first theorem applies even when Alice is allowed to use non-unitary operations to encode her messages, and also that over the entire region where any boundary saturates the bound, only unitaries can be used to encode $N$ messages (more precisely, only operations that act as unitaries on the initial entangled state; see below).

\section{Deterministic dense coding}
We begin by reviewing the initial setup for deterministic dense coding. Alice and Bob share a bipartite system, described by Hilbert space ${\cal H}_{AB}={\cal H}_A\otimes{\cal H}_B$ of dimension $d\times d$ (both subsystems have dimension $d$), in a known entangled state. This state may be written in its Schmidt decomposition \cite{NielsenChuang}, 
\begin{equation}\label{Psi0}
	\vert\Psi_0\rangle = \sum_{n=0}^{d-1} \sqrt{\lambda_n}\vert n\rangle_A\vert n\rangle_B,
\end{equation}
with $\lambda_0\ge \lambda_1\ge \cdots \ge\lambda_{d-1}$ real and positive, and $\sum_n\lambda_n=1$. Alice and Bob together choose a set of states to represent the messages she will send. When Alice uses unitary encoding, each message will be represented by a single pure state,
\begin{eqnarray}\label{Lambda}
	|\Psi_j\rangle = (U_j\otimes I_B)|\Psi_0\rangle,
\end{eqnarray}
where $I_B~(I_A)$ is the identity operator on ${\cal H}_B~({\cal H}_A)$. For the non-unitary encoding discussed in Section~\ref{sec:onlyU}, each message may be represented by more than one pure state (see below).

Alice begins by choosing a message and then encodes it by performing the corresponding local operation on her part of their shared system. She then sends system $A$ to Bob, after which he measures on the combined system $AB$ to determine Alice's message. No other communication between them is allowed. In order that Bob is able to determine with certainty the message Alice has sent, it must be that the set of states representing a given message are orthogonal to all states representing other messages. When Alice uses unitary encoding, this means that $\langle\Psi_i|\Psi_j\rangle =0~~\forall{j\ne i}$, or in terms of the $U_j$,
\begin{eqnarray}\label{Lambda}
	0 = \textrm{Tr}(U_j\Lambda U_i^\dagger),~~j\ne i,
\end{eqnarray}
a condition we will refer to as $\Lambda$-orthogonality between the unitary operators. Here, $\Lambda$ is a diagonal matrix with entries $\lambda_0,\cdots,\lambda_{d-1}$ in that order, and is in fact equal to the reduced density matrix of $|\Psi_0\rangle$,
\begin{eqnarray}
	\Lambda = \textrm{Tr}_A(|\Psi_0\rangle\langle\Psi_0|).
\end{eqnarray}
We now turn to our numerical results in $d=4$.

\section{Numerical results}\label{sec:num}
\begin{figure}[h]
\includegraphics{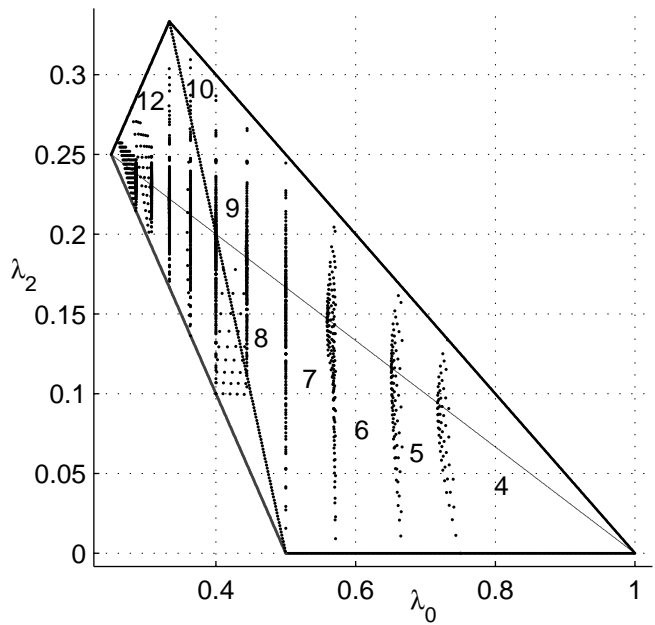}
\caption{\label{fig:d4}Phase boundaries for dense coding with a $4$-by-$4$ entangled state using unitary encoding. The $\lambda_1$ axis lies perpendicular to the page. Regions of $N$ unitary messages are labeled by the appropriate integer (except for $N=11,~13,~14$, for which there is no space, but the reader can use his or her imagination to fill in). There is no region of only $N=15=d^2-1$ unitary messages \cite{Ji}, and the ``maximally entangled" point (MES) at the far left of the plot where $\lambda_n=1/d=0.25~\forall{n}$ allows $N=d^2=16$. All boundaries to the right of regions with $N\ge 2d$ have significant portions lying at $\lambda_0=d/N$. Notice also that the boundary to the right of the region with $N=2d+1=9$ messages bends at the lower part of the plot (these data points, which in this view appear to lie within the $N=9$ region, are actually part of the boundary separating that region from the region of $N=8$) toward the triangular face corresponding to $\lambda_1=\lambda_0$ (the left-most two lines, meeting at MES, form two sides of this triangle), and that to the right of the $N=3d+1=13$ and $N=14$ regions bend at the top toward the line corresponding to $\lambda_2=\lambda_1=\lambda_0$ (the line in the upper-left of the figure extending upward from MES).}
\end{figure}

We have mapped the full phase diagram for $d=4$ and unitary encoding, the results appearing in Figure~\ref{fig:d4}. In so doing, we observed that several of the boundaries are flat planes in the three-dimensional parameter space of independent Schmidt coefficients ($\lambda_0,~\lambda_1,~\lambda_2$), located at the constant value $\lambda_0=d/N$. All boundaries for $N\ge 2d$ are flat at least over a significant portion, and several are flat over their entirety, all such flat portions lying at the value, $\lambda_0=d/N$. Those that are not entirely flat bend away from this value of $\lambda_0$ toward a particular (hyper-)plane. Specifically, for $N=2d+1=9$ and (less obviously) for $N=11$, the boundary bends toward the plane defined by the largest two Schmidt coefficients being equal, $\lambda_1=\lambda_0$; and the boundaries for $N=3d+1=13$ and $N=14$ bend toward the line with $\lambda_2=\lambda_1=\lambda_0$. We again see that the boundary for $N=d+1$ is everywhere located at $\lambda_0<d/N=0.8$, confirming the result of \cite{GerjuoyDense} that this must be so.

We note that these patterns appear for the case of $d=3$, as well (see Figure $1$ in \cite{Mozes}). The $N=2d=6$ boundary is a straight line at the constant value $\lambda_0 = d/N$, and the boundary for $N=2d+1=7$, while touching the line $\lambda_0=d/N$, tilts away from this value of $\lambda_0$ toward the line with the largest two Schmidt coefficients equal to each other, $\lambda_1=\lambda_0$. 

These observations are generalized to arbitrary dimension $d$ in theorem~\ref{th:md}, given in the next section. There, we prove that when the largest $m$ Schmidt coefficients are all equal to each other, then to have $N=md+1$ messages, it must be that $\lambda_0<d/N$.

We have also mapped the $d=4$ phase diagram when Alice is allowed to use non-unitary encoding. A very interesting observation is that it is quite common for all the operators used to encode a given set of $N$ messages to be forced to unitaries as one approaches a boundary. That is, it appears that under a wide range of circumstances, the message operators must all be unitaries right at the boundary between a region of $N$ messages and one of $N-1$ messages. There are important exceptions to this rule, however (see \cite{BeranK}, where we show there exists at least one region within which non-unitary encoding allows Alice to send strictly more messages than if she uses only unitary operations). In our second theorem, presented in Section~\ref{sec:onlyU}, we prove that all message operators must be (effectively) unitary when a boundary is located at $\lambda_0=d/N$.

\section{Shapes of boundaries}\label{sec:shapes}
We have generalized the observations described in the previous section for unitary encoding to the case of arbitrary $d$ in the following theorem.
\begin{md}\label{th:md}
When the largest $m$ Schmidt coefficients of $|\Psi_0\rangle$ are all equal to each other, $N=md+1$ unitary messages may be deterministically transmitted by dense coding only if $\lambda_0$ is strictly less than $d/N$. That is, the bound $\lambda_0=d/N$ cannot be reached  along the ``hyper-plane" defined by $\lambda_0 = \lambda_{1} = \cdots = \lambda_{m-1}$ when $N=md+1$.
\end{md}
We prove this theorem by contradiction. Thus, suppose there exist $N=md+1$ unitary messages, encoded by $U_j$ with $j=0,\cdots,md$, when the largest $m$ Schmidt coefficients are equal. We note for use below that $m<d$, since under no circumstances can $N$ exceed $d^2$, the dimension of ${\cal H}_A\otimes{\cal H}_B$. Reshape the first $m$ columns of each $U_j$ into $md$-dimensional vectors $|\phi_{0j}\rangle$ as follows: the first $d$ entries of $|\phi_{0j}\rangle$ are the $d$ entries of the first column of $U_j$ in the same order, the next $d$ entries are the entries of the second column of $U_j$, and so on. Then, normalize this vector by the factor $1/\sqrt{m}$ so that $\langle \phi_{0j}|\phi_{0j}\rangle= 1$. Next, define $d$-dimensional vectors $|\phi_{kj}\rangle$, $k=m,\cdots,d-1$, as the $k^{th}$ column of $U_j$ which, using the same normalization, will satisfy $\langle\phi_{kj}|\phi_{kj}\rangle=1/m$. 

We begin by showing that
\begin{lem1}\label{th:lem1}
If $\lambda_0 = d/N$ for $N=md+1$, then for all $i$ and $j\ne i$,
\begin{eqnarray}
	|\langle\phi_{0i}|\phi_{0j}\rangle| = \frac{1}{md}.
\end{eqnarray}
\end{lem1}
Proof: First, we show that $|\langle\phi_{0i}|\phi_{0j}\rangle| \le 1/md$. $\Lambda$-orthogonality between unitaries $U_i$ and $U_j$ with $j\ne i$ can be written as
\begin{eqnarray}
	0 = \lambda_0\langle\phi_{0i}|\phi_{0j}\rangle + \sum_{k=m}^{d-1}\lambda_k\langle\phi_{ki}|\phi_{kj}\rangle,
\end{eqnarray}
implying
\begin{eqnarray}\label{ineq}
	\lambda_0|\langle\phi_{0i}|\phi_{0j}\rangle| = \left|\sum_{k=m}^{d-1}\lambda_k\langle\phi_{ki}|\phi_{kj}\rangle\right| \le\sum_{k=m}^{d-1}\lambda_k|\langle\phi_{ki}|\phi_{kj}\rangle|\le\sum_{k=m}^{d-1}\lambda_k(\frac{1}{m}) = \frac{1}{m}(1-m\lambda_0).
\end{eqnarray}
Therefore,
\begin{eqnarray}
	|\langle\phi_{0i}|\phi_{0j}\rangle| \le \frac{1 - m\lambda_0}{m\lambda_0},
\end{eqnarray}
and setting $\lambda_0 = d/N = d/(md+1)$ gives \cite{weakNote}
\begin{eqnarray}\label{1overmd}
	|\langle\phi_{0i}|\phi_{0j}\rangle| \le \frac{1}{md}.
\end{eqnarray}

We now argue that, in fact, $|\langle\phi_{0i}|\phi_{0j}\rangle| = 1/md$ for every $j\ne i$ if $\lambda_0 = d/N$. The $N\times N$ (Hermitian) Gram matrix, $G$, of the $|\phi_{0j}\rangle$ is defined to have entries given as $G_{ij}=\langle\phi_{0i}|\phi_{0j}\rangle$. According to our choice of normalization, the diagonal elements of $G$ are all equal to unity. Since the rank of a Gram matrix is equal to the number of linearly independent vectors in the set from which it is formed, and since $G$ is formed from $md$-dimensional vectors with $md=N-1$, the rank of $G$ can be no greater than $N-1$. Therefore, $G$ must have at least one zero eigenvalue. We now turn to Gersgorin's theory on the location of eigenvalues \cite{HornJohnson}, generalized by \cite{Solovev} in the following theorem:
\begin{gersgorin}\cite{Brualdi}\label{th:gersgorin}
	Let $A = [a_{ij}]$ be an $N\times N$ matrix and let $r$ be any integer with $1\le r\le N$. Then each eigenvalue  $z$ of $A$ is either in one of the disks
\begin{eqnarray}
	\{z:|z-a_{ii}|\le S_i^{(r-1)}~~~(i = 1,2,\cdots,N)\},
\end{eqnarray}
or in one of the regions
\begin{eqnarray}
	\{z:\sum_{i\in P}|z-a_{ii}|\le \sum_{i\in P}R_i\},
\end{eqnarray}
where $S_i^{(r-1)}$ is the sum of magnitudes of the largest $r-1$ off-diagonal elements in row $i$ of $A$, $P\subseteq\{1,2,\cdots,N\}$ with $|P| = r$, and $R_i = \sum_{j\ne i}|a_{ij}|$.
\end{gersgorin}

We have at least one eigenvalue $z=0$ and every diagonal element is equal to unity for our matrix $G$, and we have just seen that every off-diagonal element of $G$ has magnitude less than or equal to $1/md$. Setting $r=N-1$, we see that $S_i^{(N-2)} \le (N-2)/md = (md-1)/md$, which is strictly less than $|z - G_{ii}| = 1$ for every $i$. Hence, the second option in the theorem must hold; that is, we require that
\begin{eqnarray}
	N-1 = \sum_{i\in P}|z-G_{ii}|\le \sum_{i\in P}\sum_{j\ne i}|\langle\phi_{0i}|\phi_{0j}\rangle| \le \frac{(N-1)^2}{md} = N-1.
\end{eqnarray}
Thus, the inequalities must hold as equalities, meaning that for at least $N-1$ of the rows of $G$, every off-diagonal entry has magnitude equal to $1/md$. However, since $G$ is Hermitian, this statement must hold also for the $N^{th}$ row, or in other words, $|\langle\phi_{0i}|\phi_{0j}\rangle|=1/md$ for all $i$ and $j\ne i$, proving the lemma. \hspace{\stretch{1}}$\blacksquare$

We can now prove our theorem.\\
Proof of theorem~\ref{th:md}: The lemma implies that the left- and right-hand sides of Eq.~(\ref{ineq}) must be equal, so the inequalities appearing there must be satisfied as equalities. Therefore, $\langle\phi_{ki}|\phi_{kj}\rangle = \xi_{ij}/m$ for all $i,j,k$, with $|\xi_{ij}|=1$ and independent of $k$. That is, for each $k$, all the $|\phi_{kj}\rangle$ are equal to each other up to a phase factor. Since we can always choose the first message to be encoded by the identity operator, whose corresponding $|\phi_{k0}\rangle$ have all entries equal to zero except the $k^{th}$ entry which is equal to $1$, then it must be that every $|\phi_{kj}\rangle$ has only its $k^{th}$ entry non-zero and equal to $\xi_{ij}=\xi_j$ (it obviously cannot depend on the arbitrary index $i$). In other words, every message operator $U_j$ is diagonal in its last $d-m$ columns (and hence also those rows), and since an overall phase factor (the $\xi_j$) in $U_j$ is irrelevant, we can set all those $d-m$ diagonal elements equal to $1$. Therefore, every $U_j$ is of the form
\begin{eqnarray}
	U_j = \left(\begin{array}{c c c c c c}
		v_{11}  & \cdots & v_{1m} &  & & \\
		\vdots & \ddots & \vdots &  & & \\
		v_{m1} & \cdots & v_{mm} &  & & \\
		& & & 1 &  & \\
		& & & & \ddots   \\
		& & & & & 1 \\
		\end{array}\right),
\end{eqnarray}
where only the non-zero entries in $U_j$ are shown, and $v = [v_{ij}]$ is an arbitrary $m\times m$ unitary matrix. There can be no more than $m^2+1$ linearly independent such matrices. However, since the $U_j$ are all pairwise ($\Lambda$) orthogonal, they must also be linearly independent. Hence, we require that $N = md+1 \le m^2+1$, which is impossible, since $m<d$. Thus, we have a contradiction, completing the proof of theorem~\ref{th:md}. \hspace{\stretch{1}}$\blacksquare$

We thus have an indication of the shapes of the boundaries in any dimension $d$, at least if Alice is restricted to unitary encoding, but what about if she can use more general operations? Is it possible for the bound to be reached when $N=md+1$ and the largest $m$ Schmidt coefficients are equal, if Alice can instead use non-unitaries to encode the messages? If so, the boundary shapes could be qualitatively different for non-unitary, as opposed to unitary, encoding. It turns out, however, that we can answer this question in the negative. In fact, we will prove an even stronger statement, which was motivated by the observation, mentioned in the introduction, that message operators tend to be forced toward unitaries as one approaches a boundary (though there are exceptions \cite{BeranK}). That is, as one moves within a region of $N$ unitary messages toward that of $N-1$ unitary messages, the $N$ (numerically generated) message operators, though allowed to be non-unitaries, generally become closer and closer to unitaries as the boundary with the $N-1$ region is approached. This leads us to the second main result of our paper, which states that all message operators must be (effectively) unitaries over those portions of a boundary that are located at the bound $\lambda_0 = d/N$.

\section{Only unitaries at the boundaries}\label{sec:onlyU}
Before we show that only unitaries work at the boundaries, it will be helpful to first review the description of protocols in which Alice uses non-unitaries \cite{ourDense,BeranK}. In this case, we will imagine that Alice brings in an ancillary system $a_j$ (of dimension $\kappa_j$) in some fixed initial state, performs unitary $U_j$ on the combination of systems $a_j$ and $A$, after which she can either measure the ancilla or throw it away. Let us suppose she measures it and obtains outcome $l$ (throwing it away does not change things in any important way). The effect on system $A$ will in general be a non-unitary operation, represented by the Kraus operator $K_{jl}$ \cite{Kraus}. As she will want to make her own choice of which message to send, rather than to allow the random outcome of her measurement on $a_j$ to determine the message, the two of them must agree that all these operations $K_{jk}$ (with fixed $j$ and $k=1, \cdots, \kappa_j$) will collectively represent message $j$. Note that unitarity of $U_j$ implies that for each $j$,
\begin{equation}\label{complete}
	I_A = \sum_{k=1}^{\kappa_j} K_{jk}^\dagger K_{jk}.
\end{equation}
In order for Bob to determine with certainty which message has been sent, the states representing message $j$,
\begin{equation}\label{prob}
	\vert\Psi_{jk}\rangle = \frac{1}{\sqrt{p_{jk}}}(K_{jk}\otimes I_B)\vert\Psi_0\rangle,
\end{equation}
with probabilities $p_{jk} = \langle\Psi_0\vert(K_{jk}^\dagger K_{jk}\otimes I_B)\vert\Psi_0\rangle$, must be orthogonal to all states corresponding to other messages, $\vert\Psi_{j^\prime k^\prime}\rangle$ for $j^\prime \ne j$.
We may think of message $j$ as being represented by the density operator,
\begin{eqnarray}\label{rhoDef}
	\rho_j = \sum_{k=1}^{\kappa_j} p_{jk}\vert\Psi_{jk}\rangle\langle\Psi_{jk}\vert,
\end{eqnarray}
and we may assume without loss of generality that for fixed $j$, states $\vert\Psi_{jk}\rangle$ are linearly independent, in which case $\kappa_j$ is known as the Kraus rank of ${\cal A}_j$ and is equal to the rank of $\rho_j$. Note that by Eqs.~(\ref{complete}) and (\ref{prob}), $\textrm{Tr}_A(\rho_j) = \textrm{Tr}_A(\vert\Psi_0\rangle\langle\Psi_0\vert) = \Lambda$ for all $j$.

We will now prove the second main result of our paper, stated in the following theorem.
\begin{onlyU}\label{th:onlyU}
	The bound, $\lambda_0 = d/N$, can only be reached by a set of message operators, each of which acts on $|\Psi_0\rangle$ as a unitary. In other words, the bound can be reached only when every message is represented by a single, pure state.
\end{onlyU}
By acting on $|\Psi_0\rangle$ ``as a unitary", we mean that for any non-unitary Kraus operator, $K_{jk}$, it must be that
\begin{eqnarray}
	|\Psi_{jk}\rangle \propto (K_{jk}\otimes I_B)|\Psi_0\rangle \propto (V_{j}\otimes I_B)|\Psi_0\rangle,
\end{eqnarray}
for some unitary $V_{j}$. We note that if $K_{jk}$ is not proportional to a unitary, then this is only possible if one or more of the Schmidt coefficients, $\lambda_n$, vanish. Then (apart from a normalization factor), $V_{j}$ and $K_{jk}$ differ only in the $n^{th}$ column (or columns, if $\lambda_n=0$ for more than one value of $n$), an irrelevant difference since changing these columns in $K_{jk}$ does not change the message, $|\Psi_{jk}\rangle$.

~

\noindent Proof: In Eq.~(24) of \cite{ourDense}, it was argued that 
\begin{equation} \label{24}
	\sum_{j=1}^N \rho_j \le \sum_{j=1}^N P_j \le I_A\otimes I_B
\end{equation}
where $P_j$ is the projector onto the  support of $\rho_j$ (${\cal M}_1\le {\cal M}_2$ is to be interpreted as saying that ${\cal M}_2 - {\cal M}_1$ is a positive semi-definite operator). Tracing both sides of this relationship over Alice's system $A$ yields
\begin{eqnarray}\label{bound}
	N\Lambda \le dI_B
\end{eqnarray}
with $\Lambda$ defined in Eq.~(\ref{Lambda}). Therefore, every $\lambda_n$ is bounded above by $d/N$, which is equivalent to the already mentioned bound, $\lambda_0 \le d/N$. Equality in this bound means, from Eq.~(\ref{24}),
\begin{equation}\label{rhoP}
	\sum_{j=1}^N \langle 0_B|\textrm{Tr}_A(\rho_j)|0_B\rangle =\sum_{j=1}^N \langle 0_B|\textrm{Tr}_A(P_j)|0_B\rangle.
\end{equation}
Write each density operator in its spectral decomposition as
\begin{equation}
	\rho_j = \sum_{k=1}^{\kappa_j} \mu_{jk} |\phi_{jk}\rangle\langle\phi_{jk}|,
\end{equation}
so that
\begin{equation}
	P_j = \sum_{k=1}^{\kappa_j} |\phi_{jk}\rangle\langle\phi_{jk}|,
\end{equation}
where $\kappa_j$ is the rank of $\rho_j$. Then,
\begin{eqnarray}\label{rho}
	\sum_{j=1}^N \langle0_B|\textrm{Tr}_A(\rho_j)|0_B\rangle = \sum_{j=1}^N \sum_{k=1}^{\kappa_j}\mu_{jk}\left|\langle0_B|\phi_{jk}\rangle\right|^2
\end{eqnarray}
and
\begin{eqnarray}\label{P}
	\sum_{j=1}^N \langle0_B|\textrm{Tr}_A(P_j)|0_B\rangle = \sum_{j=1}^N \sum_{k=1}^{\kappa_j}\left|\langle0_B|\phi_{jk}\rangle\right|^2.
\end{eqnarray}
Since $0\le\mu_{jk}\le1$, we have that each term in Eq.~(\ref{rho}) is less than or equal to the corresponding term in Eq.~(\ref{P}). Since the right-hand sides of these equations are sums of non-negative quantities, and since by Eq.~(\ref{rhoP}) the two are equal, it must be that for each $j$, one of the following two conditions holds:
\begin{eqnarray}\label{cond}
	(1)& \mu_{jk} &= ~~1~~ \textrm{(for some value of $k$, implying $\kappa_j=1$)},\nonumber \\
	\textrm{or}~~(2)& \left|\langle0_B|\phi_{jk}\rangle\right|&=~~0~~\textrm{(for all $k$ such that~} \mu_{jk} \ne 0).
\end{eqnarray}
Condition ($1$) implies that $\rho_j$ is a pure state (rank of $\rho_j$ is equal to one). If any message is not pure, then, condition ($2$) implies that $\rho_j$ must have no support on $|0_B\rangle$. This is impossible for a deterministic protocol, as the following argument shows. Since for each message $j$ of Kraus rank $\kappa_j$,
\begin{equation}
	\rho_j = \sum_{k=1}^{\kappa_j} (K_{jk}\otimes I_B)|\Psi_0\rangle\langle\Psi_0|(K_{jk}^\dagger\otimes I_B),
\end{equation}
then for condition (2) of Eq.~(\ref{cond}) to hold, it must be that for each $k$,
\begin{equation}
	0 = \langle0_B|(K_{jk}\otimes I_B)|\Psi_0\rangle=\sqrt{\lambda_0}K_{jk}|0_A\rangle.
\end{equation}
From this, we conclude that
\begin{equation}
	0 = \lambda_0\sum_{k=1}^{\kappa_j}\langle0_A|K_{jk}^\dagger K_{jk}|0_A\rangle=\lambda_0,
\end{equation}
having used Eq.~(\ref{complete}). This is a contradiction, since $\lambda_0\ge1/d$ (it is the largest of the Schmidt coefficients). Hence, for each $j$ we have that condition (1) holds, and $\rho_j$ is a pure state. This implies it was encoded by a unitary operation (more precisely, that for fixed $j$ all $K_{jk}$ are proportional to the same unitary), unless one or more of the $\lambda_n$ vanish. The case of vanishing $\lambda_n$ offers a rather trivial exception to the requirement of unitary encoding to achieve the bound, since the corresponding columns of $K_{jk}$, which are the only parts that can differ from unitarity, are then irrelevant -- they are always multiplied by zero (that is, by $\sqrt{\lambda_n}$). \hspace{\stretch{1}}$\blacksquare$

\section{Conclusions}
We have studied deterministic dense coding with partially entangled states, guided by numerical results in $d=3$ and $d=4$. We proved that when the Schmidt coefficients of the entangled state, $\lambda_n$, are such that the largest $m$ are all equal, then $N=md+1$ unitary messages can only be sent if these largest ones are strictly smaller than $d/N$. This explained the observation from numerical data that the boundaries always tilted away from $\lambda_0=d/N$ toward the hyper-planes defined by the equality of these largest Schmidt coefficients, and shows that this must also be the case in arbitrary dimensions $d$. In addition, we proved that this conclusion also holds for non-unitary encoding, and that for any number of messages $N$ and any dimension $d$, in order to saturate the bound $\lambda_0=d/N$, all messages must be (effectively) encoded by unitaries.

\begin{acknowledgments}This work has been supported in part by the National Science Foundation through Grants PHY-0456951 and PHY-0757251, and through a grant from the Research Corporation. We are grateful to Shay Mozes for sharing with us the Matlab code he used to generate the numerical results of \cite{Mozes}, and to R.B. Griffiths and his research group for numerous stimulating discussions on this topic.
\end{acknowledgments}

%\bibliography{QInfoRefs}

\begin{thebibliography}{10}

\bibitem{BennettDense}
C.~H. Bennett and S.~J. Weisner, Phys. Rev. Lett. {\bf 69},  2881  (1992).

\bibitem{Mozes}
S. Mozes, J. Oppenheim, and B. Reznik, Phys. Rev. A {\bf 71},  012311: 1
  (2005).

\bibitem{ourDense}
S. Wu, S.~M. Cohen, Y. Sun, and R.~B. Griffiths, Phys. Rev. A {\bf 73},
  042311: 1   (2006).

\bibitem{GerjuoyDense}
P.~S. Bourdon, E. Gerjuoy, J.~P. McDonald, and H.~T. Williams, Phys. Rev. A
  {\bf 77},  022305  (2008).

\bibitem{Ji}
Z. Ji {\it et~al.}, Phys. Rev. A {\bf 73},  034307: 1 – 3  (2006).

\bibitem{BeranK}
M. Beran and S.M. Cohen, arXiv:0807.4552 [quant-ph].

\bibitem{NielsenChuang}
M. Nielsen and I. Chuang, {\em Quantum Computation and Quantum Information}
  (Cambridge University Press, Cambridge, UK, 2000).

\bibitem{weakNote}
Note that these arguments work as well when, for example, $N=md+2$. However, in that case, the conclusion is that $|\langle\phi_{0i}|\phi_{0j}\rangle| \le 2/md$, which is too weak a statement to be useful.

\bibitem{HornJohnson}
R. Horn and C. Johnson, {\em Matrix Analysis} (Cambridge University Press,
  Cambridge, 1985).

\bibitem{Solovev}
V.~A. Solov'ev, Math. USSR Izvestiya {\bf 23},  545  (1984).

\bibitem{Brualdi}
R.~A. Brualdi and S. Mellendorf, The American Mathematical Monthly {\bf 101},
  975  (1994).

\bibitem{Kraus}
K. Kraus, {\em States, Effects and Operations} (Spring-Verlag, Berlin, 1983).

\end{thebibliography}
%
%\bibliographystyle{prsty}

\end{document}